\documentclass{fundam}

\usepackage[latin1]{inputenc}
\usepackage{amssymb}

\newtheorem{The}{Theorem}[section]
\newtheorem{Pro}[The]{Proposition}

\newtheorem{Lem}[The]{Lemma}
\newtheorem{Rem}[The]{Remark}
\newtheorem{Cor}[The]{Corollary}

\newcommand{\fa}{\forall}
\newcommand{\Ga}{\Gamma}
\newcommand{\Gas}{\Gamma^\star}
\newcommand{\Si}{\Sigma}
\newcommand{\Sis}{\Sigma^\star}
\newcommand{\Sio}{\Sigma^\omega}

\newcommand{\ra}{\rightarrow}
\newcommand{\hs}{\hspace{12mm}

\noi}
\newcommand{\lra}{\leftrightarrow}

\newcommand{\ite}{\item}

\newcommand{\ol}{ $\omega$-language}

\newcommand{\om}{\omega}
\newcommand{\nl}{\newline}
\newcommand{\noi}{\noindent}
\newcommand{\tla}{\twoheadleftarrow}

\newcommand{\proo}{\noi {\bf Proof.} }
\newcommand {\ep}{\hfill $\square$}

\begin{document}

\setcounter{page}{1}
\issue{(2005)}

\title{  On  Winning Conditions of High Borel \\ Complexity in Pushdown
Games }

\address{E Mail: finkel@logique.jussieu.fr}

\author{Olivier Finkel\\ Equipe de Logique Math\'ematique 
 \\ U.F.R. de Math\'ematiques, Universit\'e Paris 7 \\ 2 Place Jussieu
75251 Paris
 cedex 05, France.\\ finkel@logique.jussieu.fr }
\maketitle

\runninghead{Olivier Finkel}{ On  Winning Conditions of High Borel 
 Complexity in Pushdown
Games}

\begin{abstract}

 In a recent paper \cite{Serre04,Serre04b}
 Serre has presented some decidable winning conditions \nl 
$\Omega_{\mathcal{A}_1 \rhd \ldots \rhd \mathcal{A}_n \rhd
\mathcal{A}_{n+1}}$ of
arbitrarily high  finite Borel complexity for games on finite graphs or on
pushdown graphs.
\nl  We  answer in this paper
 several questions which were raised by Serre in \cite{Serre04,Serre04b}.
\nl We study  classes  $\mathbb{C}_n(A)$, defined in  \cite{Serre04b},
 and  show that
these classes are included in the class of non-ambiguous context free
$\om$-languages.
Moreover from the study of a  larger class  $\mathbb{C}^\lambda_n(A)$ we
infer that
the complements of languages in $\mathbb{C}_n(A)$ are also
non-ambiguous context free $\om$-languages. We conclude the study of classes
$\mathbb{C}_n(A)$ by showing that they are neither closed under union nor
under
intersection.
\nl We prove also that there exists pushdown games,  equipped with
winning conditions in the form $\Omega_{\mathcal{A}_1 \rhd \mathcal{A}_2}$,
where the winning sets are not
deterministic context free languages, giving examples of winning sets which
are
non-deterministic non-ambiguous context
free languages,  inherently ambiguous  context
free languages, or even non context free languages.

\end{abstract}


\section{Introduction}

\noi  Two-player infinite games have been much studied in set theory and in
particular in
 Descriptive Set Theory.
  Martin's Theorem states that every Gale Stewart game
 $G(A)$, where $A$ is a Borel set, is determined, i.e. that one of the
 two players has a winning strategy \cite{Kechris94}.
\nl  In Computer Science, the conditions
 of a Gale Stewart game may be seen as a specification of a reactive system,
 where the two players are respectively  a non terminating reactive
 program and  the  ``environment".
 Then the problem of the synthesis of winning strategies
is of great practical
interest for the problem of program synthesis in reactive systems.
 B\"uchi-Landweber Theorem states that in a Gale Stewart game $G(A)$,
where $A$ is a regular $\om$-language, one can decide who is the winner and
compute
a winning strategy given by a finite state transducer.
\nl In \cite{Thomas95,LescowThomas} Thomas asked for an extension of this
result to
games played on pushdown graphs. Walukiewicz firstly showed in \cite{wal}
that one can
effectively construct winning strategies in parity games played on
pushdown graphs and that these strategies can be computed by pushdown
transducers.
\nl Several authors have then studied pushdown games equipped with other
decidable
winning conditions, \cite{Cachat02, ca02d, Serre03,Gimbert}.
 Cachat, Duparc and Thomas have presented the first decidable winning
condition
at the $\Sigma_3$ level of the Borel hierarchy \cite{CDT}. Bouquet, Serre
and Walukiewicz
have studied winning conditions which are boolean combinations of a B\"uchi
condition and of
the unboundedness condition which requires the stack to be unbounded,
\cite{BSW}.
\nl Recently Serre has given a family of decidable winning conditions of
arbitrarily high
finite Borel rank
\cite{Serre04,Serre04b}. A game between two players Adam and Eve on a
pushdown graph, is
equipped with a winning condition in the form
$\Omega_{\mathcal{A}_1 \rhd \ldots \rhd \mathcal{A}_n \rhd
\mathcal{A}_{n+1}}$, where
$\mathcal{A}_1, \ldots , \mathcal{A}_n$ are deterministic pushdown automata,
the stack alphabet of  $\mathcal{A}_i$ being the input alphabet of
$\mathcal{A}_{i+1}$,  and
$\mathcal{A}_{n+1}$ is a  deterministic pushdown automaton with a B\"uchi or
a parity
acceptance condition. Then  an infinite play is won by Eve iff during this
play the
stack is {\it strictly unbounded}, that is  converges to an infinite word
$x$ and its limit
$x \in L(\mathcal{A}_1 \rhd \ldots \rhd \mathcal{A}_n \rhd
\mathcal{A}_{n+1})$, where
$L(\mathcal{A}_1 \rhd \ldots \rhd \mathcal{A}_n \rhd \mathcal{A}_{n+1})$ is
an $\om$-language
defined as follows. A word $\alpha_0$ is in
$L(\mathcal{A}_1 \rhd \ldots \rhd \mathcal{A}_n \rhd \mathcal{A}_{n+1})$
iff:
for all $1 \leq i \leq n$, when $\mathcal{A}_i$ reads $\alpha_{i-1}$ its
stack is {\it strictly
unbounded} and the limit of the stack contents is an $\om$-word $\alpha_i$;
and
$\mathcal{A}_{n+1}$ accepts $\alpha_n$. Serre proved that for these winning
conditions
one can decide the winner in a pushdown game and that the winning strategies
are effective.
\nl We solve in this paper several questions which are raised in
\cite{Serre04,Serre04b}. We
first study the classes $\mathbb{C}_n(A)$ which contain languages in the
form
$L(\mathcal{A}_1 \rhd \ldots \rhd \mathcal{A}_n \rhd \mathcal{A}_{n+1})$,
where
$A$ is the input alphabet of $\mathcal{A}_1$. We show that
these classes are included in the class of non-ambiguous context free
$\om$-languages.
Moreover from the study of a  larger class  $\mathbb{C}^\lambda_n(A)$ we
infer that
the complements of languages in $\mathbb{C}_n(A)$ are also
non-ambiguous context free $\om$-languages. We conclude the study of classes
$\mathbb{C}_n(A)$ by showing that they are neither closed under union nor
under
intersection.
\nl For all previously studied  decidable winning conditions for pushdown
games the
set of winning positions for any player had been shown to be regular.
In \cite{Serre04,Serre04b} Serre proved
that every deterministic context free
language may occur as a winning set for Eve in a pushdown game equipped with
a winning
condition in the form $\Omega_{\mathcal{B}}$, where $\mathcal{B}$ is a
deterministic
pushdown automaton.
The exact nature of  these sets remains open and the question
is raised in  \cite{Serre04,Serre04b}
whether there exists a pushdown game equipped with a winning
condition in the form
$\Omega_{\mathcal{A}_1 \rhd \ldots \rhd \mathcal{A}_n \rhd
\mathcal{A}_{n+1}}$
such that the set of winning positions for Eve is not a
 deterministic context free language.
We give a positive answer  to this question, giving examples
of winning sets which are non-deterministic non-ambiguous context
free languages, or inherently ambiguous  context
free languages, or even non context free languages.
\nl The paper is organized as follows. In section 2 we recall definitions
and results
about pushdown automata, context free ($\omega$)-languages, pushdown games,
and winning conditions in the form
$\Omega_{\mathcal{A}_1 \rhd \ldots \rhd \mathcal{A}_n \rhd
\mathcal{A}_{n+1}}$.
In section 3 are studied the classes $\mathbb{C}_n(A)$.
Results on sets of winning positions are presented
in Section 4.

\section{Recall of previous definitions and results}

\subsection{Pushdown automata}

We assume the reader to be familiar with the theory of formal
($\om$)-languages
\cite{Thomas90,Staiger97,HopcroftMotwaniUllman2001}.
We shall use usual notations of formal language theory.

\hs  When $A$ is a finite alphabet, a {\it non-empty finite word} over $A$
is any
sequence $x=a_1\ldots a_k$ , where $a_i\in A$
for $i=1,\ldots ,k$ , and  $k$ is an integer $\geq 1$. The {\it length}
 of $x$ is $k$, denoted by $|x|$.
 The {\it empty word} has no letter and is denoted by $\lambda$; its length
is $0$.
 For $x=a_1\ldots a_k$, we write $x(i)=a_i$
and $x[i]=x(1)\ldots x(i)$ for $i\leq k$ and $x[0]=\lambda$.
 $A^\star$  is the {\it set of finite words} (including the empty word) over
$A$ and
$A^+=A^\star - \{\lambda\}$.

 \hs  The {\it first infinite ordinal} is $\om$.
 An $\om$-{\it word} over $A$ is an $\om$-sequence $a_1 \ldots a_n \ldots$,
where for all
integers $ i\geq 1$, ~
$a_i \in A$.  When $\sigma$ is an $\om$-word over $A$, we write
 $\sigma =\sigma(1)\sigma(2)\ldots \sigma(n) \ldots $,  where for all $i$,~
$\sigma(i)\in A$,
and $\sigma[n]=\sigma(1)\sigma(2)\ldots \sigma(n)$  for all $n\geq 1$ and
$\sigma[0]=\lambda$.

 \hs  The {\it prefix relation} is denoted $\sqsubseteq$: a finite word $u$
is a {\it prefix}
of a finite word $v$ (respectively,  an infinite word $v$), denoted
$u\sqsubseteq v$,
 if and only if there exists a finite word $w$
(respectively,  an infinite word $w$), such that $v=u.w$.
 The {\it set of } $\om$-{\it words} over  the alphabet $A$ is denoted by
$A^\om$.
An  $\om$-{\it language} over an alphabet $A$ is a subset of  $A^\om$.

\hs   In \cite{Serre04,Serre04b} deterministic  pushdown automata
are defined with two restrictions.
It is supposed that there are no $\lambda$-transitions, i.e. the automata
are {\it real time}.
Moreover one can push at most one symbol in the pushdown stack
using a single  transition of the automaton.
\nl We  now define  pushdown automata, keeping this second restriction but
allowing
the existence of $\lambda$-transitions; and we define also the non
deterministic version of
pushdown automata.

\hs A {\it pushdown automaton} (PDA) is a 6-tuple
$\mathcal{A}=(Q, \Ga, A, \bot, q_{in}, \delta)$, where $Q$ is a finite set
of states,
$\Gamma$ is a finite pushdown alphabet, $A$ is a finite input alphabet,
$\bot$ is the bottom of stack symbol,
$q_{in} \in\ Q$ is the initial state, and $\delta$ is the transition
relation
which is a mapping from $Q \times (A \cup\{\lambda\} )\times \Ga $
to  subsets of
$$\{skip(q), pop(q), push(q, \gamma) \mid q\in Q, \gamma \in \Ga - \{\bot\}
\}$$

\noi The bottom symbol appears only at the bottom of the stack and is never
popped thus
 for all $q, q' \in Q$ and $a\in A$, it holds that $pop(q') \notin
\delta(q,  a, \bot)$.

\hs The pushdown automaton $\mathcal{A}$ is {\it deterministic}
if for all $q \in Q$, $a\in A$ and
$Z \in \Ga$, the set  $\delta(q, a, Z)$ contains at most one element;
moreover if
for some $q \in Q$ and $Z \in \Ga$,   $\delta(q, \lambda, Z)$ is non-empty
then for all
$a\in A$ the set $\delta(q, a, Z)$ is empty.

\hs If $\sigma\in\Ga^{+}$ describes the pushdown store content,
the {\it rightmost symbol} will be assumed to be {\it on ``top" of the
store}.
A configuration of the pushdown automaton $\mathcal{A}$
is a pair $(q, \sigma)$ where $q\in Q$ and
$\sigma\in\Gas$.
\nl
For $a\in A\cup\{\lambda\}$, $\sigma \in\Ga^{\star}$
and $Z\in\Ga$:
\nl if $(skip(q'))$ is in $\delta(q,a,Z)$, then we write
$a: (q, \sigma.Z)\mapsto_{\mathcal{A}} (q', \sigma.Z)$;
\nl if $(pop(q'))$ is in $\delta(q,a,Z)$, then we write
$a: (q, \sigma.Z)\mapsto_{\mathcal{A}} (q',\sigma)$;
\nl if $(push(q', \gamma))$ is in $\delta(q,a,Z)$, then we write
$a: (q, \sigma.Z)\mapsto_{\mathcal{A}} (q', \sigma.Z.\gamma)$.

\hs
$\mapsto_{\mathcal{A}}^\star$ is the transitive and reflexive closure of
$\mapsto_{\mathcal{A}}$.
(The subscript $\mathcal{A}$ will be omitted whenever the meaning remains
clear).

\hs
Let $x =a_1a_2\ldots a_n$ be a finite word over $A$.
A finite sequence of configurations $r=(q_i,\gamma_i)_{1\leq i\leq p}$ is
called
a run of $\mathcal{A}$ on $x$, starting in configuration $(q,\gamma)$, iff:
\begin{enumerate}
\ite $(q_1,\gamma_1)=(q,\gamma)$
\ite for each $i$, $1\leq i\leq (p-1)$, there exists
$b_i\in A \cup\{\lambda\}$
satisfying $b_i: (q_i,\gamma_i)\mapsto_{\mathcal{A}}(q_{i+1},\gamma_{i+1} )$
\ite
 $a_1a_2\ldots a_n =b_1b_2\ldots b_{p-1}$
\end{enumerate}

\noi
A  run $r$ of $\mathcal{A}$ on $x$, starting in configuration
$(q_{in}, \bot)$,
will be simply called ``a run of $\mathcal{A}$ on $x$".

\hs
Let $x =a_1a_2\ldots a_n\ldots$ be an $\om$-word over $A$.
An infinite sequence of configurations $r=(q_i,\gamma_i)_{i\geq1}$ is
called
a run of $\mathcal{A}$ on $x$, starting in configuration $(q,\gamma)$, iff:
\begin{enumerate}
\ite $(q_1,\gamma_1)=(q,\gamma)$

\ite for each $i\geq 1$, there exists $b_i\in A \cup\{\lambda\}$
satisfying $b_i: (q_i,\gamma_i)\mapsto_{\mathcal{A}}(q_{i+1},\gamma_{i+1} )$
\ite either ~ $a_1a_2\ldots a_n\ldots =b_1b_2\ldots b_n\ldots$
\nl or ~ $b_1b_2\ldots b_n\ldots$ is a finite prefix of ~ $a_1a_2\ldots
a_n\ldots$
\end{enumerate}

\noi The run $r$ is said to be complete when $a_1a_2\ldots a_n\ldots
=b_1b_2\ldots b_n\ldots$
\nl
A complete run $r$ of $\mathcal{A}$ on $x$, starting in configuration
$(q_{in}, \bot)$,
will be simply called ``a run of $\mathcal{A}$ on $x$".

\hs If the pushdown automaton  $\mathcal{A}$ is equipped with a set of final
states
$F \subseteq Q$,
\nl the finitary language {\it accepted by } $(\mathcal{A}, F)$ is :
$$L^f(\mathcal{A}, F)= \{ x \in A^\star \mid \mbox{ there exists a run }
r=(q_i,\gamma_i)_{1\leq i\leq p}
\mbox{ 
of } \mathcal{A}  \mbox{ on } x  \mbox{  such that }  q_p\in F \}$$
\noi The class  $CFL$ of  {\it  context free languages} 
 is the class of finitary languages which are 
accepted by pushdown automata by final states. 
\nl Notice that
other accepting conditions by PDA have been shown to be equivalent to the
acceptance 
condition by final states. Let us cite, \cite{ABB96}:
(a) acceptance  by empty storage, (b) acceptance by final states and empty storage, 
(c) acceptance by topmost stack letter, 
(d) acceptance  by final states and topmost stack letter.
\nl  The class $DCFL$ of {\it deterministic 
context free languages} is the class of finitary languages which are 
accepted by deterministic  pushdown automata (DPDA) by final states. 
\nl Notice that for DPDA, acceptance by final states is not 
equivalent to acceptance by empty storage: this is due to the fact that 
a language accepted by a DPDA by empty storage 
must be {\it prefix-free} while this is not necessary in the case of 
 acceptance by final states \cite{ABB96}. 

\hs The $\om$-language {\it B\"uchi accepted by} $(\mathcal{A}, F)$ is :
$$L(\mathcal{A}, F)= \{ x \in A^\om \mid \mbox{ there exists a  run } r \mbox{ 
of } \mathcal{A}  \mbox{ on } x  \mbox{  such that }  In(r) \cap F \neq\emptyset \}$$ 
\noi where 
$In(r)$ is  the set of all states
entered infinitely
often during run $r$.

\hs If instead the pushdown automaton  $\mathcal{A}$ is equipped with a set of accepting sets 
of  states $\mathcal{F} \subseteq 2^Q$, 
the $\om$-language {\it Muller accepted by} $(\mathcal{A}, \mathcal{F})$ is :
$$L(\mathcal{A}, \mathcal{F})= \{ x \in A^\om \mid \mbox{ there exists a  run } r \mbox{ 
of } \mathcal{A}  \mbox{ on } x  \mbox{  such that }  In(r) \in  \mathcal{F} \}$$ 
\noi The class $CFL_\om$ of  
{\it context free } $\om$-{\it languages} is the class of $\om$-languages which are 
B\"uchi or Muller accepted by pushdown automata. 

\hs Another usual acceptance condition for $\om$-words is the parity condition.
In that case a pushdown automaton $\mathcal{A}=(Q, \Ga, A, \bot, q_{in},
\delta)$
is equipped with a function $col$ from $Q$ to a finite set of colors $C
\subset \mathbb{N}$.
The $\om$-language accepted by  $(\mathcal{A}, col)$  is:
$$L(\mathcal{A}, col)= \{ x \in A^\om \mid \mbox{ there exists a  run } r
\mbox{ 
of } \mathcal{A}  \mbox{ on } x  \mbox{  such that }  sc(r) \mbox{ is even } \}$$ 
\noi where $sc(r)$ is the smallest color appearing infinitely often in the run $r$. 
\nl It is easy to see that a B\"uchi acceptance condition can be expressed as a parity 
acceptance condition which itself can be expressed as a Muller condition. 
\nl Thus the class of $\om$-languages which are 
accepted by pushdown automata with a parity acceptance condition is still the class 
$CFL_\om$. 

\hs Consider now {\it deterministic} pushdown automata.   
If $\mathcal{A}$ is a deterministic pushdown automaton, then for every $\sigma\in A^\om$,
 there exists at most one run $r$ of $\mathcal{A}$ on $\sigma$ determined 
by the starting configuration.
The pushdown automaton  has the continuity property iff for every $\sigma \in A^\om$, 
there exists a 
 unique run of $\mathcal{A}$ on $\sigma$ and this run is complete. It is shown in \cite{CG78} 
that each $\om$-language accepted by a deterministic B\"uchi (respectively,  Muller) 
pushdown automaton
can be accepted by a deterministic B\"uchi (respectively,  Muller) 
pushdown automaton  with the continuity property. 
The same proof works in the case of deterministic pushdown automata with parity 
acceptance condition. 

\hs The class of $\om$-languages accepted by deterministic B\"uchi 
pushdown automata is a strict subclass of the class $DCFL_\om$ of $\om$-languages 
accepted by deterministic pushdown automata with a Muller condition. 
\nl One can easily show that $DCFL_\om$ is also the class 
of $\om$-languages 
accepted by DPDA with a parity 
acceptance condition. 
\nl Each $\om$
-language in $DCFL_\om$ can be accepted  by a deterministic pushdown
automaton {\it having the  continuity property }with parity (or Muller)
acceptance condition.
One can then show that the class $DCFL_\om$ is closed under complementation.

\hs The notion of ambiguity for context free $\om$-languages has been
firstly studied in
\cite{fin03b}.
 A context free $\om$-language is non ambiguous iff it is accepted by a
B\"uchi or Muller pushdown automaton such that every $\om$-word on the input
alphabet
has at most one accepting run. Notice that we consider here that two runs
are equal iff they
go through the same infinite sequence of configurations {\it and }
$\lambda$-transitions  occur
at the same steps of the computations.
\nl The class $NA-CFL_\om$ is the class of non ambiguous context free
$\om$-languages.
\nl The inclusion $DCFL_\om  \subseteq  NA-CFL_\om$ will be useful in the
sequel.
\noi We shall denote $Co-NA-CFL_\om$ the class of complements of non
ambiguous
context free $\om$-languages.

\subsection{Pushdown games}

\noi Recall first that a {\it pushdown process} may be viewed as
a PDA without input alphabet and initial state.
 A pushdown process is a 4-tuple
$\mathcal{P}=(Q, \Ga,  \bot,  \delta)$, where $Q$ is a finite set of states,
$\Gamma$ is a finite pushdown alphabet,
$\bot$ is the bottom of stack symbol,
and $\delta$ is the transition relation
which is a mapping from $Q \times \Ga $
to  subsets of
$$\{skip(q), pop(q), push(q, \gamma) \mid q\in Q, \gamma \in \Ga - \{\bot\}
\}$$

\hs  Configurations of a pushdown process are defined as for PDA.
A configuration of the pushdown process  $\mathcal{P}$
is a pair $(q, \sigma)$ where $q\in Q$ and
$\sigma\in\Gas$.

\hs To a pushdown process $\mathcal{P}=(Q, \Ga,  \bot,  \delta)$ is
naturally associated
a pushdown graph $G=(V, \ra)$ which is a directed graph. The set of vertices
$V$
is the set of configurations of $\mathcal{P}$. The edge relation $\ra$ is
defined as follows:
$(q, \sigma)\ra (q', \sigma')$
iff the configuration $(q', \sigma')$  can be  reached
in one transition of  $\mathcal{P}$ from the configuration $(q, \sigma)$.

\hs We shall consider in the sequel infinite games between two players named
Eve  and Adam
on such pushdown graphs.
\nl So we shall assume that the set $Q$ of states of a pushdown process is
partitioned in
two sets $Q_E$ and $Q_A$.
A configuration $(q, \sigma)$ is in $V_E$ iff $q$ is in $Q_E$ and it is
 in $V_A$ iff $q$ is in $Q_A$ so $(V_E, V_A)$ is a partition of the set of
configurations $V$.

\hs The game graph $(V_E, V_A, \ra)$ is called a {\it pushdown game graph}.
\nl A play from a vertex $v_1$ of this graph is defined as follows. If
$v_1 \in V_E$, Eve chooses a vertex $v_2$ such that $v_1 \ra v_2$; otherwise
Adam chooses
such a vertex. If there is no such vertex $v_2$ the play stops. Otherwise
the play may continue.
If $v_2 \in V_E$, Eve chooses a vertex $v_3$ such that $v_2 \ra v_3$;
otherwise Adam chooses
such a vertex. If there is no such vertex $v_3$ the play stops. Otherwise
the play continues in the
same way.
So a  play starting from the vertex $v_1$ is a  {\it finite or infinite}
sequence
of vertices $v_1v_2v_3 \ldots$ such that for all $i$ $v_i \ra v_{i+1}$.
 We may assume, as in \cite{Serre04,Serre04b},
that in fact all plays are infinite.

\hs A {\it winning condition} for Eve is a set $\Omega \subseteq V^\om$.
An infinite two-player pushdown game is a 4-tuple $(V_E, V_A, \ra, \Omega)$,
where
$(V_E, V_A, \ra)$ is a pushdown game graph and $\Omega \subseteq V^\om$ is
a winning condition for Eve.
\nl In a pushdown game equipped with the winning condition $\Omega$,
Eve wins a play $v_1v_2v_3 \ldots$ iff  $v_1v_2v_3 \ldots \in \Omega$.
\nl A {\it strategy} for Eve is a partial function $f : V^\star.V_E \ra V$
such
that,  for all $x \in V^\star$ and $ v\in V_E$, $v \ra f(x.v)$.
\nl Eve uses the strategy $f$ in a play $v_1v_2v_3 \ldots$ iff
for all $v_i \in V_E$, $v_{i+1}=f(v_1v_2\ldots v_i)$.
\nl A strategy $f$ is a {\it winning strategy}  for Eve from
some position $v_1$ iff Eve wins all plays starting from $v_1$ and
during which she uses the strategy $f$.
\nl  A vertex $v\in V$ is a {\it winning position} for Eve iff she has a
winning
strategy from it.
\nl The notions of winning
strategy and winning position are defined for the other player Adam in a
similar way.
\nl The set of winning positions for Eve and Adam will be respectively
denoted by
$W_E$ and $W_A$.

\subsection{Winning condition
$\Omega_{\mathcal{A}_1 \rhd \ldots \rhd \mathcal{A}_n \rhd
\mathcal{A}_{n+1}}$ }

\noi
We first recall the definition of $\om$-languages
in the form $L(\mathcal{A}_1 \rhd \ldots \rhd \mathcal{A}_{n} \rhd
\mathcal{A}_{n+1})$ which
are used in \cite{Serre04,Serre04b} to define the winning conditions
$\Omega_{\mathcal{A}_1 \rhd \ldots \rhd \mathcal{A}_n \rhd
\mathcal{A}_{n+1}}$.

\hs We shall need  the notion of limit of an infinite sequence of finite
words over some finite alphabet $A$.
\nl Let then $(\beta_n)_{n\geq 0}$ be an infinite sequence of words $\beta_n
\in A^\star$.
The finite {\it or} infinite word $\lim_{n\in\om} \beta_n$ is
determined by the set of its (finite) prefixes:
for all  $v$ in   $A^\star$,
\nl $v\sqsubseteq \lim_{n\in\om} \beta_n \lra  \exists n \fa p\geq n\quad
\beta_p[|v|]=v$.

\hs Let now  $\mathcal{A}=(Q, \Ga, A, \bot, q_{in}, \delta)$ be a pushdown
automaton
reading words over the alphabet $A$ and let $\alpha \in A^\om$. The pushdown
stack of
$\mathcal{A}$ is said to be {\it strictly unbounded} during a run
$r=(q_i,\gamma_i)_{i\geq1}$ of $\mathcal{A}$ on  $\alpha$
iff $\lim_{n \geq 1} \gamma_n$ is infinite.

\hs We define now $\om$-languages
 $L(\mathcal{A}_1 \rhd \ldots \rhd \mathcal{A}_n \rhd \mathcal{A}_{n+1})$
in a slightly more general case than in \cite{Serre04b}, because this will
be useful in the
next section.
 Notice that in \cite{Serre04b},
these $\om$-languages are only  defined in the  case where
$\mathcal{A}_1,  \ldots , \mathcal{A}_n$, are {\it real-time }
deterministic pushdown automata, and
$\mathcal{A}_{n+1}$ is a {\it real-time } deterministic
pushdown automaton equipped with a parity or a  B\"uchi acceptance
condition.

\hs  Let $n$ be an integer $\geq 0$ and $\mathcal{A}_1$, $\mathcal{A}_2$,
\ldots
$\mathcal{A}_n$, be some deterministic pushdown automata (in the case $n=0$
there are
not any such automata).
\nl Let
$(\mathcal{A}_{n+1}, \mathcal{C})$ be a
pushdown automaton equipped with a B\"uchi or a parity acceptance condition.
\nl The input alphabet of $\mathcal{A}_1$ is denoted $A$ and we assume that,
for each integer $i \in [1, n]$, the input
alphabet of  $\mathcal{A}_{i+1}$ is the stack alphabet of $\mathcal{A}_i$.

\hs  We define inductively
the $\om$-language $L(\mathcal{A}_1 \rhd \ldots \rhd \mathcal{A}_n \rhd
\mathcal{A}_{n+1})
\subseteq A^\om$ by:

\begin{enumerate}
\ite If $n=0$, $L(\mathcal{A}_1 \rhd \ldots \rhd \mathcal{A}_n \rhd
\mathcal{A}_{n+1}) =
L(\mathcal{A}_{n+1}, \mathcal{C})$ is the $\om$-language accepted by
$\mathcal{A}_{n+1}$ with
acceptance condition $\mathcal{C}$.
\ite If $n>0$, $L(\mathcal{A}_1 \rhd \ldots \rhd \mathcal{A}_n \rhd
\mathcal{A}_{n+1})$ is the
set of $\om$-words $\alpha \in A^\om$ such that:
\begin{itemize}
\ite When $\mathcal{A}_1$ reads $\alpha$, the stack of $\mathcal{A}_1$ is
strictly unbounded
hence the sequence of stack contents has an infinite  limit $\alpha_1$.
\ite $\alpha_1 \in  L(\mathcal{A}_2 \rhd \ldots \rhd \mathcal{A}_n \rhd
\mathcal{A}_{n+1})$.
\end{itemize}
\end{enumerate}

\hs  Let now $(V_E, V_A, \ra)$ be a pushdown game graph associated with a
pushdown process
$\mathcal{P}$. An infinite play  $v_1v_2v_3 \ldots$,  where $v_i=(q_i,
\gamma_i)$, is in
the set $\Omega_{\mathcal{A}_1 \rhd \ldots \rhd \mathcal{A}_n \rhd
\mathcal{A}_{n+1}}$
iff:
\begin{enumerate}
\ite
The pushdown stack of
$\mathcal{P}$ is  {\it strictly unbounded} during the play, i.e.
$\lim_{n \geq 1} \gamma_n$ is infinite, and
\ite
$\lim_{n \geq 1} \gamma_n \in
L(\mathcal{A}_1 \rhd \ldots \rhd \mathcal{A}_n \rhd \mathcal{A}_{n+1})$.
\end{enumerate}

\section{Classes $\mathbb{C}_n(A)$  }

\subsection{Classes $\mathbb{C}_n(A)$  and context free $\om$-languages}

\noi For each integer $n\geq 0$ and each
 finite alphabet $A$ the class $\mathbb{C}_n(A)$ is defined
in \cite{Serre04b} as the class of $\om$-languages in the form
$L(\mathcal{A}_1 \rhd \ldots \rhd \mathcal{A}_n \rhd \mathcal{A}_{n+1})$,
where
$\mathcal{A}_1,  \ldots , \mathcal{A}_n$, are {\it real-time }
deterministic pushdown automata,  the input
alphabet of $\mathcal{A}_1$ being $A$, and
$\mathcal{A}_{n+1}$ is a {\it real-time } deterministic
pushdown automaton equipped with a parity acceptance
condition. It is easy to see that we obtain the same class $\mathbb{C}_n(A)$
if we
restrict the definition to the case of {\it real-time } deterministic
pushdown automata  $\mathcal{A}_1,  \ldots , \mathcal{A}_n,
\mathcal{A}_{n+1}$,
having the {\it continuity property}.
\nl We shall denote  $\mathbb{C}^\lambda_n(A)$ the class obtained in the
same way except
that the deterministic pushdown automata
 $\mathcal{A}_1,  \ldots , \mathcal{A}_n, \mathcal{A}_{n+1}$,
having still the  continuity property,
may have $\lambda$-transitions, i.e. may be non real time.

\hs In the sequel of this paper when we consider languages in the form
$L(\mathcal{A}_1 \rhd \ldots \rhd \mathcal{A}_n \rhd \mathcal{A}_{n+1})$, we
shall always
implicitely assume  that the pushdown automata
 $\mathcal{A}_1,  \ldots , \mathcal{A}_n, \mathcal{A}_{n+1}$,
 have the {\it continuity property}, and that,  for each integer
$i \in [1, n]$, the input
alphabet of  $\mathcal{A}_{i+1}$ is the stack alphabet of $\mathcal{A}_i$.

\hs In order to prove that  classes  $\mathbb{C}_n(A)$,
$\mathbb{C}^\lambda_n(A)$,
  are included in the class of context free
$\om$-languages we first state the following lemma.

\begin{Lem}\label{lem1}
Let  $\mathcal{A}_1=(Q_1, \Ga_1, A_1, \bot_1, q_0^1, \delta_1)$ be a
deterministic pushdown automaton and
$\mathcal{A}_2=(Q_2, \Ga_2, \Ga_1, \bot_2, q_0^2, \delta_2)$ be a
pushdown automaton equipped with a set of final states $F \subseteq Q_2$.
Then the $\om$-language $L(\mathcal{A}_1  \rhd \mathcal{A}_2)$  is a context
free
$\om$-language.
\end{Lem}

\proo
  Let  $\mathcal{A}_1=(Q_1, \Ga_1, A_1, \bot_1, q_0^1, \delta_1)$ be a
deterministic pushdown automaton and
$\mathcal{A}_2=(Q_2, \Ga_2, \Ga_1,\nl  \bot_2, q_0^2, \delta_2)$ be a
pushdown automaton equipped with a set of final states $F \subseteq Q_2$.

\hs Recall that an $\om$-word $\alpha \in A_1^\om$ is in $L(\mathcal{A}_1
\rhd \mathcal{A}_2)$
iff:
\begin{itemize}
\ite When $\mathcal{A}_1$ reads $\alpha$, the stack of $\mathcal{A}_1$ is
strictly unbounded
hence the sequence of stack contents has an infinite  limit $\alpha_1$.
\ite $\alpha_1 \in  L(\mathcal{A}_2, F)$.
\end{itemize}

\noi We can decompose the reading of an $\om$-word
$\alpha \in L(\mathcal{A}_1  \rhd \mathcal{A}_2)$ by the pushdown automaton
$\mathcal{A}_1$ in the following way.
\nl When reading $\alpha$, $\mathcal{A}_1$ goes through the infinite
sequence of configurations
$(q_i, \gamma_i)_{i\geq 1}$. The infinite sequence of stack contents
$(\gamma_i)_{i\geq 1}$
has limit $\alpha_1$ thus
for each integer $j\geq 1$, there is a smallest integer $n_j$ such that,
for all integers
$i\geq n_j$,
$\alpha_1[j]=\gamma_i[j]$.

\hs The word $\alpha$ can then be decomposed in the form
$$\alpha = \sigma_1.\sigma_2 \ldots \sigma_n \ldots$$
\noi where for all integers $j\geq 1$, $\sigma_j \in A_1^\star$ and
$$\sigma_j:  (q_{n_j}, \alpha_1[j]) \mapsto_{\mathcal{A}_1}^\star
 (q_{n_{j+1}}, \alpha_1[j+1]) = (q_{n_{j+1}}, \alpha_1[j].\alpha_1(j+1)) $$

\noi Notice that $n_1=1$, $q_1=q_0^1$ and $\alpha_1[1]=\bot_1$ hence
$\sigma_1:  (q_{0}^1, \bot_1) \mapsto_{\mathcal{A}_1}^\star
 (q_{n_{2}}, \alpha_1[2])$.

\hs Let now, for each    $q, q' \in Q_1 \mbox{ and } a, b \in \Ga_1$,
the language  $\mathcal{L}_{( q,  q',  a,  b )}$ be the set of words
$ \sigma \in A_1^\star$
such that:
$\sigma : (q, a) \mapsto_{\mathcal{A}_1}^\star (q', a.b)$.
\noi This language of finite words over $A_1$ is accepted by the pushdown
automaton
$\mathcal{A}_1$ with
the following modifications: the initial configuration is $(q, a)$
and the acceptance is by final state $q'$ {\it and } by final stack content
$a.b$. It is easy to see that this language is also accepted by a
deterministic pushdown  automaton by final states so it is in the class
$DCFL$.

\hs Then each word $\sigma_j$ belongs to the deterministic context free
language
$$\mathcal{L}_{( q_{n_j},  q_{n_{j+1}},  \alpha_1(j),  \alpha_1(j+1) )} =
   \{ \sigma \in A_1^\star \mid \sigma :
((q_{n_j}, \alpha_1(j)) \mapsto_{\mathcal{A}_1}^\star
(q_{n_{j+1}}, \alpha_1(j).\alpha_1(j+1))\}$$

\hs In order to  describe the $\om$-language $L(\mathcal{A}_1  \rhd
\mathcal{A}_2)$ from
the $\om$-language $L(\mathcal{A}_2, F)$ and the deterministic
context free languages $\mathcal{L}_{( q,  q',  a,  b )}$, for $q, q' \in
Q_1$ and
$a, b \in \Ga_1$, we now recall the notion of substitution.

\hs A {\it substitution}  is  a mapping
$f: \Si\ra 2^{\Ga^\star}$, where $\Si$ and $\Ga$ are two finite alphabets.
If $\Si=\{a_1, \ldots ,a_n\}$, then
for all integers $i\in [1;n]$, $f(a_i)=L_i$ is a finitary language
over the alphabet  $\Ga$.
\nl Now this mapping is extended in the usual manner to finite words: for all letters 
$a_{i_1}, \ldots,  a_{i_n} \in \Si$, 
$f(a_{i_1} \ldots a_{i_n})= f(a_{i_1}) \ldots f(a_{i_n})$,
and to finitary languages $L\subseteq \Sis$:
$f(L)=\cup_{x\in L} f(x)$.
\nl If for each letter $a\in \Si$,  the language  $f(a)$ does not
contain the empty word, then the substitution is said to be $\lambda$-free
and
the mapping $f$ may be extended to $\om$-words:
$$f(x(1)\ldots x(n)\ldots )= \{u_1\ldots u_n \ldots  \mid \fa i \geq 1 \quad
u_i\in f(x(i))\}$$
\noi and to \ol s $L\subseteq \Sio$ by setting  $f(L)=\cup_{x\in L}
f(x)\subseteq \Gamma^\om$.
\nl If the substitution is not $\lambda$-free we can define $f(L)$ in the
same way for
$L\subseteq \Sio$ but
this time $f(L) \subseteq \Gas \cup \Ga^\om$, i.e. $f(L)$ may contain finite
{\it or } infinite
words.
\nl The substitution $f$ is said to be a context free substitution iff for
all $a\in \Si$ the
finitary language $f(a)$ is context free.
\nl Recall that  Cohen and Gold proved  in \cite{CG} that if $L$ is a
context free
$\om$-language and $f$ is a context free substitution then
$f(L) \cap \Gas$ and $f(L) \cap \Ga^\om$ are
context free.

\hs We define now a new alphabet
$$\Delta = \{ L(q, q', a, b) \mid q, q' \in Q_1 \mbox{ and } a, b \in \Ga_1
\}$$

\noi and we consider  the substitution
$h: \Ga_1 \ra  2^\Delta$ defined, for all $b \in \Ga_1$, by:
$$h(b) = \{ L(q, q', a, b) \mid q, q' \in Q_1 \mbox{ and } a \in \Ga_1 \}$$

\noi Applying this substitution to the
$\om$-language $L(\mathcal{A}_2, F) \subseteq \Ga_1^\om$, we get
$h( L(\mathcal{A}_2, F) )$.
The substitution $h$ is $\lambda$-free thus $h( L(\mathcal{A}_2, F) )$ is a
$\om$-language
over $\Delta$. Moreover for each $b \in \Ga_1$ the set $h(b)$ is finite
hence context free.
Thus $h( L(\mathcal{A}_2, F) )\subseteq \Delta^\om$ is a context free
$\om$-language because  $L(\mathcal{A}_2, F)$
is      a context free $\om$-language and     the substitution
$h$ is a context free substitution.

\hs Let now $R \subseteq \Delta^\om$ be the $\om$-language defined as
follows. An $\om$-word
$x\in R$ has its first letter in the set
 $\{ L(q_0^1, q',\bot_1, b) \mid q' \in Q_1 \mbox{ and } b \in \Ga_1 \}$,
and  each letter
$L(q, q', a, b)$, for $q, q' \in Q_1 \mbox{ and } a, b \in \Ga_1$,  in  $x$
is followed by a letter in the set
$\{ L(q', q'', b, c) \mid  q'' \in Q_1 \mbox{ and } c \in \Ga_1 \}$.
\nl The $\om$-language $R$ is regular thus
$h( L(\mathcal{A}_2, F) ) \cap R \subseteq \Delta^\om$ is a
context free $\om$-language because  the class $CFL_\om$ is closed under
intersection with
regular $\om$-languages \cite{CG}.

\hs Consider now the substitution $\Theta : \Delta \ra 2^{A_1^\star}$
defined,
for all letters
$ L(q, q', a, b) \in \Delta$, by
$\Theta ( L(q, q', a, b) ) = \mathcal{L}_{( q,  q',  a,  b )}$.
The   substitution    $\Theta$ is  context free   thus
$$  \Theta [ h( L(\mathcal{A}_2, F) ) \cap R ] \cap A_1^\om $$
\noi is a context free $\om$-language and so is
$ \bot_1.( ~ \Theta [ h( L(\mathcal{A}_2, F) ) \cap R ] \cap A_1^\om ~)$.
 By construction this $\om$-language is
$L(\mathcal{A}_1  \rhd \mathcal{A}_2)$.
\ep

\hs We can in fact obtain a refined result if the language
$L(\mathcal{A}_2, F)$ is non ambiguous.

\begin{Lem}\label{lem2}
Let  $\mathcal{A}_1=(Q_1, \Ga_1, A_1, \bot_1, q_0^1, \delta_1)$ be a
deterministic pushdown automaton and
$\mathcal{A}_2=(Q_2, \Ga_2, \Ga_1, \bot_2, q_0^2, \delta_2)$
be a  pushdown automaton
equipped with a set of final states $F \subseteq Q_2$.
If the $\om$-language $L(\mathcal{A}_2, F)$ is non ambiguous then
$L(\mathcal{A}_1  \rhd \mathcal{A}_2) \in NA-CFL_\om$.
\end{Lem}

\proo Let  $\mathcal{A}_1=(Q_1, \Ga_1, A_1, \bot_1, q_0^1, \delta_1)$ be a
deterministic pushdown automaton and
$\mathcal{A}_2=(Q_2, \Ga_2, \Ga_1, \nl \bot_2, q_0^2, \delta_2)$ be a
pushdown automaton equipped with a set of final states $F \subseteq Q_2$.

\hs We assume that $L(\mathcal{A}_2, F)$ is non ambiguous so we can assume,
without loss
of generality, that the pushdown automaton $\mathcal{A}_2$ itself is non
ambiguous.

\hs We are going to explain informally the construction of a non ambiguous
B\"uchi
pushdown automaton $\mathcal{A}$
accepting the $\om$-language $L(\mathcal{A}_1  \rhd \mathcal{A}_2)$.

\hs We refer now to the proof of the preceding lemma.
 We have considered  the reading of an $\om$-word
$\alpha \in L(\mathcal{A}_1  \rhd \mathcal{A}_2)$ by
$\mathcal{A}_1$, and we have shown that the word $\alpha$ can then be
decomposed in the form
$$\alpha = \sigma_1.\sigma_2 \ldots \sigma_n \ldots$$
\noi where for all integers $j\geq 1$, $\sigma_j$
belongs to the deterministic context free language
$$\mathcal{L}_{( q_{n_j},  q_{n_{j+1}},  \alpha_1(j),  \alpha_1(j+1) )} =
   \{ \sigma \in A_1^\star \mid \sigma :
((q_{n_j}, \alpha_1(j)) \mapsto_{\mathcal{A}_1}^\star
(q_{n_{j+1}}, \alpha_1(j).\alpha_1(j+1))\}$$
\noi We can see that the integers $n_j$ were defined in a unique way.
However there may exist several  decompositions
of the $\om$-word $\alpha$ into words of languages
$\mathcal{L}_{( q,  q',  a,  b )}$.
\nl In order to ensure a unique decomposition we are going to slightly
modify the
definition of these languages.
\nl For each    $q, q' \in Q_1 \mbox{ and } a, b \in \Ga_1$,  the
language         $\mathcal{U}_{( q,  q',  a,  b )}$ is the set of words $
\sigma \in A_1^\star$
such that:
\begin{enumerate}
\ite[(a)]
$\sigma : (q, a) \mapsto_{\mathcal{A}_1}^\star (q', a.b)$ and
\ite[(b)]  If for some $\sigma' \sqsubset \sigma$ and $s \in Q$,
$\sigma' : (q, a) \mapsto_{\mathcal{A}_1}^\star (s, a.b)$ then there is a
word
$u \in A_1^\star$ and a state $t \in Q$, such  that $\sigma'.u \sqsubseteq
\sigma$ and
 $u : (s, a.b) \mapsto_{\mathcal{A}_1}^\star (t, a)$.
\ite[(c)] If there is a run $(q_i, \gamma_i)_{1\leq i\leq p}$ of
$\mathcal{A}_1$ on
$\sigma$ such that $(q_1, \gamma_1)=(q, a)$ and $(q_p, \gamma_p)=(s, a.b)$
for some
 $s\in Q$, $s\neq q'$, then either there is an integer $p'<p$ such that
$(q_i, \gamma_i)_{1\leq i\leq p'}$ is a run of $\mathcal{A}_1$ on
$\sigma$ and $(q_{p'}, \gamma_{p'})=(q', a.b)$
or it holds that $\lambda : (s, a.b) \mapsto_{\mathcal{A}_1}^\star (s', a)$
for some
$s'\in Q$ and
$\lambda : (s', a) \mapsto_{\mathcal{A}_1}^\star (q', a.b)$.
\end{enumerate}

\noi It is easy to see that the languages $\mathcal{U}_{( q,  q',  a,  b )}$
are also
in the class $DCFL$ and that, for
 each    $q, q' \in Q_1 \mbox{ and } a, b \in \Ga_1$,  it holds that
$\mathcal{U}_{( q,  q',  a,  b )} \subseteq \mathcal{L}_{( q,  q',  a,
 b )}$.
\nl We can see that,   in the above decomposition
$\alpha = \sigma_1.\sigma_2 \ldots \sigma_n \ldots$  of the $\om$-word
$\alpha$,
for all integers $j\geq 1$, the word   $\sigma_j$
belongs in fact to the deterministic context free language
$\mathcal{U}_{( q_{n_j},  q_{n_{j+1}},  \alpha_1(j),  \alpha_1(j+1) )}$.

\hs  The rest of the proof of Lemma \ref{lem1} can be pursued, replacing
languages
$\mathcal{L}_{( q,  q',  a,  b )}$ by languages
$\mathcal{U}_{( q,  q',  a,  b )}$.

\hs But now we have a unique decomposition of $\alpha$ in the form
$$\alpha = \sigma'_1.\sigma'_2 \ldots \sigma'_n \ldots$$
\noi where for all integers $j \geq 1$, the word $\sigma'_j$ belongs to
some language $\mathcal{U}_{( s_j,  t_j,  a_j,  b_j )}$ satisfying:
(1) $s_1=q_0^1$, $a_1=\bot_1$, (2) for all integers $j \geq 1$,
$t_j=s_{j+1}$ and
$b_j=a_{j+1}$.

\hs This unique decomposition is crucial in the construction of the  non
ambiguous
B\"uchi PDA $\mathcal{A}$ accepting $L(\mathcal{A}_1  \rhd \mathcal{A}_2)$.
We shall explain informally the behaviour of this automaton.
\nl  For each    $q, q' \in Q_1 \mbox{ and } a, b \in \Ga_1$,  the
language         $\mathcal{U}_{( q,  q',  a,  b )}$ is  accepted by a
deterministic
pushdown automaton $\mathcal{B}^{( q,  q',  a,  b )}$ whose stack alphabet
is denoted
$\Ga^{( q,  q',  a,  b )}$. We can assume that all these alphabets are
disjoint and that
they are also disjoint from $\Ga_1$, the stack alphabet of $\mathcal{A}_1$.
 The stack alphabet of $\mathcal{A}$ will be
$$\Ga^\mathcal{A}=\Ga_1 \cup
\bigcup_{q, q' \in Q_1 ~and ~ a, b \in \Ga_1} \Ga^{( q,  q',  a,  b )}$$

\noi When reading an $\om$-word
$\alpha \in L(\mathcal{A}_1  \rhd \mathcal{A}_2)$ the pushdown automaton
$\mathcal{A}$ will
guess, using the non determinism,  the {\bf  unique} decomposition of
$\alpha$
in the form
$$\alpha = \sigma'_1.\sigma'_2 \ldots \sigma'_n \ldots$$
\noi where for all integers $j \geq 1$, the word $\sigma'_j$ belongs to
some language $\mathcal{U}_{( s_j,  t_j,  a_j,  b_j )}$ satisfying:
(1) $s_1=q_0^1$, $a_1=\bot_1$, (2) for all integers $j \geq 1$,
$t_j=s_{j+1}$ and
$b_j=a_{j+1}$.
\nl In addition $\mathcal{A}$ will simulate the reading of
the $\om$-word $\alpha_1=a_1a_2a_3\ldots$
by the PDA $\mathcal{A}_2$.

\hs During a run of   $\mathcal{A}$ the stack content is always a word
in the form $\bot.u.v$ where $\bot$ is the bottom symbol of $\mathcal{A}$,
$u \in (\Ga_1-\{\bot\})^\star$ and $v$ is in $( \Ga^{( q,  q',  a,
 b )} )^\star$ for
some  $q, q' \in Q_1$ and $ a, b \in \Ga_1$.

\hs After having read the initial segment $\sigma'_1.\sigma'_2 \ldots
\sigma'_j$ of $\alpha$,
the content of the stack of $\mathcal{A}$ is equal to the content of the
stack
of $\mathcal{A}_2$ after
having read $a_1a_2\ldots a_j$.
\nl Then $\mathcal{A}$ guesses that the next word in the  decomposition of
$\alpha$
 belongs to some $\mathcal{U}_{( s_{j+1},  t_{j+1},  a_{j+1},  b_{j+1} )}$.
It uses the
stack alphabet $\Ga^{( s_{j+1},  t_{j+1},  a_{j+1},  b_{j+1} )}$ on the top
of the stack
to simulate the reading of $\sigma'_{j+1}$
by $\mathcal{B}^{( s_{j+1},  t_{j+1},  a_{j+1},  b_{j+1} )}$.
Then when it has guessed that it has completely read the word
$\sigma'_{j+1}$, it erases
letters of $\Ga^{( s_{j+1},  t_{j+1},  a_{j+1},  b_{j+1} )}$ from the stack,
and simulates
the reading of the letter $a_{j+1}$ by $\mathcal{A}_2$, and so on.
\nl A B\"uchi acceptance condition is then used to simulate the acceptance
of $\alpha_1$ by $\mathcal{A}_2$.

\hs
The B\"uchi PDA $(\mathcal{A}_2, F)$ is non ambiguous and the above cited
decomposition of
$\alpha$ is unique so there is a unique accepting run of the B\"uchi PDA
$\mathcal{A}$ on $\alpha$.
\nl Finally we have proved that $L(\mathcal{A}_1  \rhd \mathcal{A}_2) \in
NA-CFL_\om$.
\ep

\begin{Pro}\label{pro}
Let $n$ be an integer $\geq 1$, $\mathcal{A}_1$, $\mathcal{A}_2$, \ldots
$\mathcal{A}_n$, be some deterministic pushdown automata and
$(\mathcal{A}_{n+1}, \mathcal{C})$ be a
pushdown automaton equipped with a B\"uchi acceptance condition.
 The input alphabet of $\mathcal{A}_1$ is denoted $A$ and we assume that,
for each integer $i \in [1, n]$, the input
alphabet of  $\mathcal{A}_{i+1}$ is the stack alphabet of $\mathcal{A}_i$.
Then   $L(\mathcal{A}_1 \rhd \ldots \rhd \mathcal{A}_n \rhd
\mathcal{A}_{n+1})
\in CFL_\om$.  Moreover if $L(\mathcal{A}_{n+1}, \mathcal{C})$ is non
ambiguous then
$L(\mathcal{A}_1 \rhd \ldots \rhd \mathcal{A}_n \rhd \mathcal{A}_{n+1}) \in
NA-CFL_\om$.
\end{Pro}

\proo We reason by induction on the integer $n$.
\nl  For $n=1$ the result is stated in the above Lemmas \ref{lem1} and
\ref{lem2}.

\hs Assume now that the result is true for some integer $n\geq 1$.
\nl Let $\mathcal{A}_1$, $\mathcal{A}_2$, \ldots
$\mathcal{A}_n$, $\mathcal{A}_{n+1}$, be some deterministic pushdown
automata and
$(\mathcal{A}_{n+2}, \mathcal{C})$ be a
pushdown automaton equipped with a B\"uchi acceptance condition such that
the language
$L(\mathcal{A}_1 \rhd \ldots \rhd \mathcal{A}_{n+1} \rhd \mathcal{A}_{n+2})
\subseteq A^\om$ is well defined.
\nl By induction hypothesis the language
$L(\mathcal{A}_2\rhd \ldots \rhd \mathcal{A}_{n+1} \rhd \mathcal{A}_{n+2})$
is
a context free $\om$-language accepted by a B\"uchi pushdown automaton
$(\mathcal{A}, F)$.
\nl But by definition of the language
$L(\mathcal{A}_1 \rhd \ldots \rhd \mathcal{A}_{n+1} \rhd \mathcal{A}_{n+2})$
it holds that

$$L(\mathcal{A}_1 \rhd \ldots \rhd \mathcal{A}_{n+1} \rhd \mathcal{A}_{n+2})
=
L(\mathcal{A}_1 \rhd \mathcal{A})$$
\noi thus Lemma \ref{lem1} implies that
$L(\mathcal{A}_1 \rhd \ldots \rhd \mathcal{A}_{n+1} \rhd \mathcal{A}_{n+2})
\in CFL_\om$.

\hs Assume now that $L(\mathcal{A}_{n+1}, \mathcal{C})$ is non ambiguous.
Reasoning as above
but applying Lemma \ref{lem2} instead of Lemma \ref{lem1} we infer that
$L(\mathcal{A}_1 \rhd \ldots \rhd \mathcal{A}_{n+1} \rhd \mathcal{A}_{n+2})$
is in
$NA-CFL_\om$.
\ep

\hs In particular,  Proposition \ref{pro} implies  the following result.

\begin{Cor}\label{cor1}
For each integer $n\geq 0$, the following inclusions hold:
$$\mathbb{C}_n(A) \subseteq \mathbb{C}^\lambda_n(A) \subseteq NA-CFL_\om$$
\end{Cor}

\noi We shall later get a stronger result (see Corollary \ref{cor3-8})
from the  study of closure properties of  classes
$\mathbb{C}_n(A)$, $ \mathbb{C}^\lambda_n(A)$.

\subsection{Closure properties of classes $\mathbb{C}_n(A)$, $
\mathbb{C}^\lambda_n(A)$}

We first state the following lemma.

\begin{Lem}\label{lemcomplement}
The class $ \mathbb{C}^\lambda_1(A)$ is closed under
complementation.
\end{Lem}

\proo
 Let  $\mathcal{A}_1=(Q_1, \Ga_1, A_1, \bot_1, q_0^1, \delta_1)$ be a
deterministic pushdown automaton and
$(\mathcal{A}_2=(Q_2, \Ga_2, \Ga_1, \nl  \bot_2, q_0^2, \delta_2), col_2)$ be a
deterministic pushdown automaton equipped with a parity acceptance
condition.
\nl  Recall that an $\om$-word $\alpha \in A_1^\om$ is in $L(\mathcal{A}_1
\rhd \mathcal{A}_2)$
iff: when $\mathcal{A}_1$ reads $\alpha$, the stack of $\mathcal{A}_1$ is
strictly unbounded
and the sequence of stack contents has an infinite  limit
$\alpha_1 \in  L(\mathcal{A}_2, col_2)$.

\hs  Thus an $\om$-word $\alpha \in A_1^\om$ is in the complement of
 $L(\mathcal{A}_1  \rhd \mathcal{A}_2)$
iff one of  the two following conditions holds:
\begin{itemize}
\ite[(1)] When $\mathcal{A}_1$ reads $\alpha$, the stack of $\mathcal{A}_1$
is strictly unbounded
and the  limit $\alpha_1$ of stack contents is in the complement of $
L(\mathcal{A}_2, col_2)$.
\ite[(2)] When $\mathcal{A}_1$ reads $\alpha$, the stack of $\mathcal{A}_1$
is {\it not strictly unbounded}.
\end{itemize}

\noi The class $DCFL_\om$ is closed under complementation thus the
complement of $ L(\mathcal{A}_2, col_2)$ is equal to $ L(\mathcal{A}_3,
col_3)$, for some
 deterministic
pushdown automaton $\mathcal{A}_3$ equipped with a parity acceptance
condition.
\nl The language $L(\mathcal{A}_1  \rhd \mathcal{A}_3)$
is the set of $\om$-words $\alpha \in A_1^\om$ such that,
when $\mathcal{A}_1$ reads $\alpha$,
the stack of $\mathcal{A}_1$ is strictly unbounded
and   the  limit $\alpha_1$ of stack contents is in
$L(\mathcal{A}_3, col_3)$.  So we see that, in order to get
the complement of  $L(\mathcal{A}_1  \rhd \mathcal{A}_2)$
we have to add to $L(\mathcal{A}_1  \rhd \mathcal{A}_3)$ the set $B$ of
all $\om$-words $\alpha \in A_1^\om$ such that,
when $\mathcal{A}_1$ reads $\alpha$,
the stack of $\mathcal{A}_1$ is {\it not strictly unbounded}.
\nl To do this we are going first to modify the automaton  $\mathcal{A}_1$
in such a way
that, when reading $\om$-words in $B$, the stack will be {\it strictly
unbounded}.

\hs We now explain informally the behaviour of the new pushdown automaton
$\mathcal{A}'_1$. The stack  alphabet of $\mathcal{A}'_1$ is $\Ga_1 \cup
\Ga'_1$, where
$\Ga'_1=\{\gamma' \mid \gamma\in \Ga_1\}$ is just a copy of $\Ga_1$, such
that
$\Ga_1 \cap \Ga'_1 = \emptyset$.
\nl The essential idea is that $\mathcal{A}'_1$ will simulate
$\mathcal{A}_1$ but it has
the additional following behaviour. Using $\lambda$-transitions it pushes in
the stack
letters of $\Gamma'_1$, always keeping the information about the content of
the stack of
$\mathcal{A}_1$.
\nl More precisely, if at some step while reading  an $\om$-word $\alpha \in
A_1^\om$ by
$\mathcal{A}_1$ the stack content is a finite word $\gamma = \gamma_1,
\gamma_2, \ldots
\gamma_j$, where each $\gamma_i$ is a letter of $\Ga_1$, then the
corresponding
stack content of $\mathcal{A}'_1$ will be in the form
$\gamma_1.\gamma_1^{'n_1} \gamma_2.\gamma_2^{'n_2} \ldots
\gamma_j.\gamma_j^{'n_j}$,
where $n_1, n_2, \ldots , n_j$, are positive integers.
\nl If when $\mathcal{A}_1$ reads $\alpha$ the stack is strictly unbounded
and the limit of the stack contents  is an $\om$-word $\alpha_1$, then when
$\mathcal{A}'_1$ reads the same word $\alpha$ its  stack will be  also
strictly unbounded
and the limit of the stack contents will be  an $\om$-word $\alpha'_1$.
Moreover it will
hold that
$(\alpha'_1 / \Ga'_1)=\alpha_1$, where $(\alpha'_1 / \Ga'_1)$ is the word
$\alpha'_1$
from which are removed all letters in $\Ga'_1$.
\nl On the other hand if when $\mathcal{A}_1$ reads $\alpha$ the stack {\it
is not
strictly unbounded}, the limit of the stack contents being a finite word
$\alpha_1$,
then when $\mathcal{A}'_1$ reads the same word $\alpha$ its  stack
{\it will be   strictly unbounded} and
its limit will be   an $\om$-word $\alpha'_1$ such that $(\alpha'_1 /
\Ga'_1)=\alpha_1$.

\hs Notice that the stack content of $\mathcal{A}'_1$ will always be in the
form
$\bot_1.(\bot'_1)^\star$ or
$u.Z.(Z')^n$ for some $u\in \bot_1.(\Ga_1 \cup \Ga'_1)^\star$, $Z \in
\Ga_1$,
$Z'$ being the copy of $Z$ in  $\Ga'_1$, and $n\geq 0$ being an integer.

\hs The behaviour of the deterministic pushdown automaton $\mathcal{A}'_1$,
reading an
$\om$-word, will be the same as the   behaviour of $\mathcal{A}_1$
but with the following modifications.

\begin{enumerate}
\ite[(a)]
Between any two transitions of $\mathcal{A}_1$ is added a
$\lambda$-transition of
$\mathcal{A}'_1$ which simply pushes in the stack,
when the topmost stack letter of $\mathcal{A}'_1$
 is $Z \in \Ga_1$ or $Z' \in  \Ga'_1$,
an additional letter $Z'$.

\ite[(b)] Assume now that at some step of the reading of $\alpha$ by
$\mathcal{A}'_1$ and
$\mathcal{A}_1$,  and after the execution of a $\lambda$-transition as
explained
in above item $(a)$, the topmost stack letter of $\mathcal{A}'_1$
is some letter $Z' \in \Ga'_1$.
Recall that the
stack content of $\mathcal{A}'_1$ will be
in the form $\bot_1.(\bot'_1)^n$ (if $Z'=\bot'_1$)
or $u.Z.(Z')^n$ for some $u\in \bot_1.(\Ga_1 \cup \Ga'_1)^\star$, $Z \in
\Ga_1$,
$Z'$ being the copy of $Z$ in  $\Ga'_1$, and $n\geq 1$.
\nl Notice that the corresponding stack content of $\mathcal{A}_1$ will be
$\bot_1$ or
$(u / \Ga'_1).Z$.
\nl Suppose now that  $\mathcal{A}_1$
reads a letter $a \in A_1$ or executes a $\lambda$-transition.
\nl If it
pushes letter $T$ in the stack then  $\mathcal{A}'_1$
would push the same letter $T$ in its stack.
\nl If  $\mathcal{A}_1$
would skip (its  topmost stack letter being $Z$), then  $\mathcal{A}'_1$
also skips.
\nl But if $\mathcal{A}_1$,  reading  the
letter $a \in A_1$ or executing  a $\lambda$-transition, the topmost stack
letter being $Z$,
would pop the  letter $Z$, then $\mathcal{A}'_1$
pops the whole segment $Z.(Z')^n$ at the top of the stack, using
$\lambda$-transitions.
\end{enumerate}

\noi Notice that we do not detail here the set of states of
$\mathcal{A}'_1$. It contains
the set of states $Q_1$ of $\mathcal{A}_1$ and is sufficiently enriched, to
achieve
the goal of simulating the behaviour of $\mathcal{A}_1$, adding  the
modifications cited
above.

\hs Assume now that when  $\mathcal{A}_1$ reads $\alpha$ its  stack is
strictly unbounded
and the limit of the stack contents  is an $\om$-word $\alpha_1$. Then when
$\mathcal{A}'_1$ reads the same word $\alpha$ its  stack is also strictly
unbounded
and the limit of the stack contents will be  an $\om$-word $\alpha'_1$ such
that
$(\alpha'_1 / \Ga'_1)=\alpha_1$.
\nl On the other hand if when $\mathcal{A}_1$ reads $\alpha$ the stack {\it
is not
strictly unbounded}, then the  limit of its  stack contents is a finite word
$\alpha_1=\alpha_1(1).\alpha_1(2) \ldots \alpha_1(|\alpha_1|)$.
\nl In that case   when $\mathcal{A}'_1$ reads the same word $\alpha$ its
stack
{\it will be   strictly unbounded} and
its limit will be   an $\om$-word $\alpha'_1$ in the form
$$\alpha'_1=\alpha_1(1).(\alpha_1(1)')^{n_1}.\alpha_1(2).(\alpha_1(2)')^{n_2
} \ldots
(\alpha_1(|\alpha_1|-1)')^{n_{|\alpha_1|-1}}.(\alpha_1(|\alpha_1|).(\alpha_1
(|\alpha_1|)')^{\om}$$
\noi for some integers $n_1, n_2, \ldots ,n_{|\alpha_1|-1}$.
 In particular it will hold that $(\alpha'_1 / \Ga'_1)=\alpha_1$.

\hs It is now easy to modify the pushdown automaton $\mathcal{A}_3$ in such
a way
that we obtain a  deterministic pushdown automaton $\mathcal{A}'_3$ equipped
with
parity acceptance condition $col'_3$ such that the input alphabet of
$\mathcal{A}'_3$ is
    $\Ga_1 \cup \Ga'_1$,  and an
$\om$-word $\alpha'_1 \in (\Ga_1 \cup \Ga'_1)^\om$ is in $L(\mathcal{A}'_3,
col'_3)$
 iff [  $(\alpha'_1 / \Ga'_1)$ is a finite word or $(\alpha'_1 / \Ga'_1)$ is
infinite and is in
$L(\mathcal{A}_3, col_3)$ ].
\nl  Thus the $\om$-language $L(\mathcal{A}'_1  \rhd \mathcal{A}'_3)$ is the
complement
of $L(\mathcal{A}_1  \rhd \mathcal{A}_2)$ and this ends the proof.
\ep

\begin{Pro}\label{complement}
For each integer $n\geq 0$, the class $ \mathbb{C}^\lambda_n(A)$ is closed
under
complementation.
\end{Pro}

\proo  We  now reason by induction on the integer $n \geq 0$.

\hs For $n=0$, $ \mathbb{C}^\lambda_0(A)=DCFL_\om$ is known to be closed
under complementation
\cite{Staiger97}.

\hs For $n=1$, $ \mathbb{C}^\lambda_1(A)$ is closed under
complementation by  Lemma \ref{lemcomplement}.

\hs Assume now that  we have proved that for every positive integer $k\leq
n$ the class
$\mathbb{C}^\lambda_k(A)$ is closed under
complementation.

\hs
Let $\mathcal{A}_1$, $\mathcal{A}_2$, \ldots
$\mathcal{A}_n$, $\mathcal{A}_{n+1}$, be some deterministic pushdown
automata and
$(\mathcal{A}_{n+2}, col)$ be a deterministic
pushdown automaton equipped with a parity acceptance condition such that the
language
$L(\mathcal{A}_1 \rhd \ldots \rhd \mathcal{A}_{n+1} \rhd \mathcal{A}_{n+2})
\subseteq A_1^\om$ is well defined.
\nl An $\om$-word $\alpha \in A_1^\om$ is in the complement of
$L(\mathcal{A}_1 \rhd \ldots \rhd \mathcal{A}_{n+1} \rhd \mathcal{A}_{n+2})$
iff one the two following conditions holds:
\begin{itemize}
\ite[(1)] When $\mathcal{A}_1$ reads $\alpha$, the stack of $\mathcal{A}_1$
is strictly unbounded
and the  limit $\alpha_1$ of stack contents is in the complement of
$L(\mathcal{A}_2 \rhd \ldots \rhd \mathcal{A}_{n+1} \rhd \mathcal{A}_{n+2})$
\ite[(2)] When $\mathcal{A}_1$ reads $\alpha$, the stack of $\mathcal{A}_1$
is {\it not strictly unbounded}.
\end{itemize}

\noi By induction hypothesis the complement of the $\om$-language
$L(\mathcal{A}_2 \rhd \ldots \rhd \mathcal{A}_{n+1} \rhd \mathcal{A}_{n+2})$
is in $\mathbb{C}^\lambda_n(A)$ so it is in the form
$L(\mathcal{A}'_2 \rhd \ldots \rhd \mathcal{A}'_{n+1} \rhd
\mathcal{A}'_{n+2})$.
\nl We can do similar modifications as in the case $n=1$, replacing
$\mathcal{A}_1$,
whose stack alphabet is $\Ga_1$, by another deterministic pushdown automaton
$\mathcal{A}'_1$,
whose alphabet is $\Ga_1 \cup \Ga'_1$ where $\Ga'_1$ is a copy of  $\Ga_1$.
\nl If when  $\mathcal{A}_1$ reads $\alpha$ the limit of its stack
contents is a finite or infinite word $\alpha_1$ then when
$\mathcal{A}'_1$ reads the same word $\alpha$ the limit of its stack
contents is an $\om$-word $\alpha'_1$ such that
$(\alpha'_1 / \Ga'_1)=\alpha_1$.
\nl It is now easy to modify the language
$L(\mathcal{A}'_2 \rhd \ldots \rhd \mathcal{A}'_{n+1} \rhd
\mathcal{A}'_{n+2})$ in such
a way that we get a language
$L(\mathcal{A}''_2 \rhd \ldots \rhd \mathcal{A}''_{n+1} \rhd
\mathcal{A}''_{n+2})$ of
$\om$-words over $\Ga_1 \cup \Ga'_1$ containing an $\om$-word $\alpha'_1$ if
and only if:
either $(\alpha'_1 / \Ga'_1)$ is a finite word or $(\alpha'_1 / \Ga'_1)$
belongs to the
$\om$-language $L(\mathcal{A}'_2 \rhd \ldots \rhd \mathcal{A}'_{n+1} \rhd
\mathcal{A}'_{n+2})$.

\hs Thus it holds that
$L(\mathcal{A}'_1 \rhd\mathcal{A}''_2 \rhd \ldots
\rhd \mathcal{A}''_{n+1} \rhd \mathcal{A}''_{n+2})$ is the complement of
$L(\mathcal{A}_1 \rhd \ldots \rhd \mathcal{A}_{n+1} \rhd
\mathcal{A}_{n+2})$.
\ep

\begin{Rem}
In \cite{Serre04,Serre04b} Serre defined  winning conditions
$\Omega_{\mathcal{A}_1 \rhd \ldots \rhd \mathcal{A}_n \rhd
\mathcal{A}_{n+1}}$
for pushdown games using
languages in classes $ \mathbb{C}_n(A)$.
He then showed that these winning conditions lead to decision procedures to
decide the
winner in pushdown games.
The question now naturally arises whether
the proofs can be extended to winning conditions defined in the same way
from classes $\mathbb{C}^\lambda_n(A)$.
Then the closure under complementation of these classes
would be  relevant from a game point of view.
On the other hand this closure property  provides also some more
information about classes  $ \mathbb{C}_n(A)$, given by next corollary,
which is already important from a game point of view.
\end{Rem}

\begin{Cor}\label{cor3-8}
For each integer $n\geq 0$, the following inclusions hold:
$$\mathbb{C}_n(A) \subseteq \mathbb{C}^\lambda_n(A)
\subseteq NA-CFL_\om ~\bigcap ~Co-NA-CFL_\om$$
\end{Cor}

\proo It follows directly from Corollary \ref{cor1} and Proposition
\ref{complement}.
\ep

\hs We  now prove  that the classes $ \mathbb{C}_n(A)$,
$\mathbb{C}^\lambda_n(A)$, are
not closed under other boolean operations.

\begin{Pro}
For each integer $n\geq 0$, the classes $ \mathbb{C}_n(A)$ and
$\mathbb{C}^\lambda_n(A)$
are neither closed under union nor under intersection.
\end{Pro}

\proo Notice first that  for  each integer $n\geq 0$,
$ \mathbb{C}_n(A) \subseteq \mathbb{C}_{n+1}(A)$ and
$ \mathbb{C}^\lambda_n(A) \subseteq \mathbb{C}^\lambda_{n+1}(A)$.

\hs  The $\om$-languages
$L_1=\{ a^n.b^m.c^p.d^\om \mid n, m, p \geq 1 \mbox{ and } n=m \}$
and $L_2=\{ a^n.b^m.c^p.d^\om \mid n, m, p \geq 1 \mbox{ and } m=p \}$, over
the alphabet
$A=\{a, b, c, d\}$,
are in $DCFL_\om$ and they are in all classes
  $ \mathbb{C}_n(A)$ and  $\mathbb{C}^\lambda_n(A)$.
But their intersection is
$L_1 \cap L_2=\{ a^n.b^n.c^n.d^\om \mid n \geq 1  \}$. This $\om$-language
 is not context free
because the finitary language $\{ a^n.b^n.c^n \mid n \geq 1  \}$ is not
context free
\cite{ABB96} and an $\om$-language in the form $L.d^\om$, with $L\subseteq
\{a,b,c\}^\star$,
is context free iff the finitary language $L$ is context free \cite{CG}.
Thus $L_1 \cap L_2$ cannot be in any class
$ \mathbb{C}_n(A)$ and  $\mathbb{C}^\lambda_n(A)$ because these classes are
included in $CFL_\om$.

\hs  On the other hand consider the $\om$-languages
$L_3=\{ a^n.b^m.c^p.d^\om \mid n, m, p \geq 1 \mbox{ and } n \neq m \}$
and $L_4=\{ a^n.b^m.c^p.d^\om \mid n, m, p \geq 1 \mbox{ and } m \neq p \}$.
These $\om$-languages are in $DCFL_\om$ and in every class $
\mathbb{C}_n(A)$ or
$\mathbb{C}^\lambda_n(A)$.
If the language $L_3 \cup L_4$ was in some class $ \mathbb{C}_n(A)$ or
$\mathbb{C}^\lambda_n(A)$, then  by Proposition \ref{complement}
its complement $L_5$ would be also in $\mathbb{C}^\lambda_n(A)$ and it would
be a
context free $\om$-language. This would imply that  $L_5 \cap
a^+.b^+.c^+.d^\om$
is context free because the class $CFL_\om$ is closed under intersection
with
regular $\om$-languages.
But $L_5 \cap a^+.b^+.c^+.d^\om=\{ a^n.b^n.c^n.d^\om \mid n \geq 1  \}$ is
not context free
thus for each integer $n\geq 0$, the classes $ \mathbb{C}_n(A)$,
$\mathbb{C}^\lambda_n(A)$ are
not closed under union.

\hs Notice that  the union $\cup_{n\geq 0}\mathbb{C}^\lambda_n(A)$ is also
neither closed under intersection nor under union.
\ep

\section{Winning sets in a pushdown game}

\noi Recall that it is proved in \cite{Serre04} that every deterministic
context free
language may occur as a winning set for Eve in a pushdown game equipped with
a winning
condition in the form $\Omega_{\mathcal{B}}$, where $\mathcal{B}$ is a
deterministic
pushdown automaton.
\nl Serre asked also whether there exists a pushdown game equipped with a
winning
condition in the form
$\Omega_{\mathcal{A}_1 \rhd \ldots \rhd \mathcal{A}_n \rhd
\mathcal{A}_{n+1}}$
such that the set of winning positions for Eve is not a
 deterministic context free language.
\nl We are going to prove  in this section that such pushdown games exist,
giving examples of winning sets which are non-deterministic non-ambiguous
context
free languages, or inherently ambiguous  context
free languages, or even non context free languages.
\nl
The exact form of the winning sets remains open.  
 Serre conjectured in \cite{SerrePhd} that one could prove that,
for $n\geq 0$,
the winning sets for Eve in pushdown games equipped with a winning
condition in the form
$\Omega_{\mathcal{A}_1 \rhd \ldots \rhd \mathcal{A}_n \rhd
\mathcal{A}_{n+1}}$,  form
a class of languages at level $n$,
and that
for $n=0$ the winning sets could be deterministic context free
languages.
\nl So we think that, in order to better understand what is the exact form of
the
winning sets, it is useful
 to see different examples of
winning sets of different complexities, and not only of the greatest complexity 
we have got, i.e. a  non context free language. 
\nl Moreover the techniques,  involving Duparc's eraser operator, used to
prove Proposition
\ref{prows} below, are interesting by their own and are useful
to understand how the games go on.

\hs In order to present the first example we begin by recalling
the operation $x \ra x^\tla $ which has been introduced by Duparc
in his study of the Wadge hierarchy \cite{Duparc01}, where it works also on
infinite words,
and is also considered by Serre in \cite{Serre04}.

\hs For a finite word $u\in  (\Si \cup \{\tla\})^\star$,    where $\Si$ is a
finite alphabet,
the finite word $u^\tla$ is inductively defined by:

\hs  $\lambda^\tla =\lambda$,
\nl and for a finite word $u\in (\Si \cup \{ \tla\})^\star$:
\nl $(u.c)^\tla=u^\tla.c$, if $c\in \Si$,
\nl $(u.\tla)^\tla =u^\tla$  with its last letter removed if $|u^\tla|>0$,
\nl i.e. $(u.\tla)^\tla =u^\tla(1).u^\tla(2)\ldots u^\tla(|u^\tla|-1)$  if
$|u^\tla|>0$,
\nl $(u.\tla)^\tla=\lambda$  if $|u^\tla|=0$,

\hs Notice that for  $x \in (\Si \cup \{ \tla\})^{\star}$, $x^\tla$ denotes
the string $x$,
once every $\tla$ occuring in $x$, used as an eraser,
has been ``evaluated" to the back space operation,
proceeding from left to right inside $x$. In other words $x^\tla = x$ from
which every
interval of the form $``c\tla "$ ($c \in \Si$) is removed.
\nl For a language $V \subseteq \Si^\star$ we set
$V^\sim = \{ x \in (\Si \cup \{ \tla\})^\star \mid x^\tla \in  V \}$.

\begin{Lem}\label{lemna}
Let $L = \{a^n.b^n \mid  n\geq 1\}$. Then $L^\sim$ is a non ambiguous
context free language
which can not be accepted by any {\it deterministic
pushdown automaton}.
\end{Lem}

\proo Let  $L$ be  the  context free language
$\{a^n.b^n \mid  n\geq 1\}$.
 The language $L$ is a deterministic, hence non ambiguous, context free
language. Thus by Theorem 6.16 of \cite{fin03b} the language
$L^\sim$ is a non ambiguous context free language.
\nl It remains to show that $L^\sim$ can not be accepted by any {\it
deterministic
pushdown automaton}.

\hs The idea of the proof is essentially the same as in the proof that the
context free language $\{a^n.b^n \mid n\geq 1\} ~\cup ~\{a^n.b^{2n} \mid
n\geq 1\}$
can not be accepted by any {\it deterministic
pushdown automaton}. It can be found in \cite[Proof of Proposition
5.3]{ABB96} or in
\cite[Exercise 6.4.4 page 251]{HopcroftMotwaniUllman2001}.
\nl Towards a contradiction assume that the language $L^\sim$ is accepted by
a
deterministic pushdown automaton $\mathcal{A}$.
All words $a^n.b^n$, for $n\geq 1$, are in the language $L^\sim$. Then one
could show that
there exists a pair $(n, k)$, with $n, k > 0$, such that the accepting
configurations
of $\mathcal{A}$ reading $a^n.b^n$ or $a^{n+k}.b^{n+k}$ are the same.
Consider now the word $a^n.b^n.\tla^{2n}.a.b$. It belongs to $L^\sim$ and
the valid
computation of $\mathcal{A}$ reading $a^n.b^n$ should be the beginning of
the valid
computation of $\mathcal{A}$ reading $a^n.b^n.\tla^{2n}.a.b$. Thus the
pushdown automaton
$\mathcal{A}$ would also accepts $a^{n+k}.b^{n+k}.\tla^{2n}.a.b$ which is
clearly not in
$L^\sim$.
\ep

\begin{Lem}\label{lemna2}
Let $L \subseteq \Sis$ be a deterministic context free language. Then there
exists a pushdown process $\mathcal{P}=(Q, \Ga,  \bot,  \delta)$, a
partition
$Q = Q_E \cup Q_A$,  two deterministic  pushdown automata $\mathcal{A}_1,
\mathcal{A}_2$,
and a state  $q\in Q$ such that, in the induced pushdown game
equipped with the winning condition
$\Omega_{\mathcal{A}_1 \rhd \mathcal{A}_2}$,
one has  $\{u \mid (q, u) \in W_E \} = L^\sim$.
\end{Lem}

\proo Let $\mathcal{P}=(\{p,q\}, \Ga=\Si \cup \{\bot,   \tla, \#\},  \bot,
\delta)$
be a pushdown process
where $\delta$ is defined by: $push(p, \#) \in \delta(q, c)$ for all letters
$c\in \Si \cup \{\bot, \tla\}$ and $push(p, \#) \in \delta(p, \#)$.
\nl So the pushdown process $\mathcal{P}$ is deterministic and its behaviour
is very similar to the behaviour of the pushdown process given in the proof
of Proposition 42 of \cite{Serre04b}. It can only push the letter $\#$ on the
top
of a given configuration.
\nl $Q = Q_E \cup Q_A$ is any partition of  $Q$.
\nl For each configuration $(q, u.c)$, for $c\in \Si \cup \{\bot,  \tla\}$
and $u\in \Gas$,
there is a unique infinite play starting from $(q, u.c)$, during which the
pushdown stack of
$\mathcal{P}$ is strictly unbounded, and the limit of the stack contents is
$u.c.\#^\om$.

\hs The deterministic pushdown automaton $\mathcal{A}_1$ reads words over
the alphabet
$\Ga=\Si \cup \{\bot, \tla, \#\}$ and its stack alphabet is  $\Ga_1=\Si \cup
\{\bot_1\}$.
Its behaviour is described as follows:

\hs Consider first the reading of an $\om$-word in the form $\bot.u.\#^\om$,
where $u \in (\Si \cup \{\tla\})^\star$.
\nl After having read the bottom symbol $\bot$, the content
of its stack is still $\bot_1$. Then when the pushdown automaton
$\mathcal{A}_1$ reads  a
letter $c\in \Si$ it pushes the same letter in the stack.
But if $\mathcal{A}_1$ reads the symbol $\tla$ and the topmost stack symbol
is not $\bot_1$ (so it is in $\Si$) then it pops the letter at the top
of its stack.

\hs So, after having read the initial segment $\bot.u$ of  $\bot.u.\#^\om$,
the stack content
of $\mathcal{A}_1$ is $\bot_1.u^\tla$.
 Next  the PDA  $\mathcal{A}_1$ pushes a letter $\#$ in the stack
for each letter $\#$ read.
 Thus, when $\mathcal{A}_1$ reads the $\om$-word $\bot.u.\#^\om$,  its stack
is
strictly unbounded and the limit of the stack contents is
$\bot_1.u^\tla.\#^\om$.

\hs In addition, it is easy to ensure that, when $\mathcal{A}_1$ reads an
$\om$-word
which is not in $\bot.(\Si \cup \{ \tla\})^\star.\#^\om \cup  \bot.(\Si \cup
\{ \tla\})^\om$,
then its stack {\it is not strictly
unbounded}. If there is a letter $\bot$  after the first letter of the word
or if
$\mathcal{A}_1$ reads a letter in $\Si \cup \{\tla\}$ after some letter
$\#$, then the stack
content remains undefinitely unchanged.

\hs On the other hand, $\mathcal{A}_2$ is  a deterministic pushdown
automaton equipped
with a parity acceptance condition which accepts the $\om$-language
$\bot_1.L.\#^\om $.

\noi  Consider now a given configuration $(q, \bot.u)$ of the pushdown
process $\mathcal{P}$
for some $u \in (\Si \cup \{\tla, \#\})^\star$, the last letter of $u$ being
not $\#$.
 There is a unique infinite play starting from this position. The stack of
$\mathcal{P}$ is
strictly unbounded during this play and the limit of stack contents is
$\bot.u.\#^\om$.
\nl When $\mathcal{A}_1$ reads the $\om$-word $\bot.u.\#^\om$ its stack is
strictly unbounded iff  $u\in (\Si \cup \{ \tla\})^\star$
 and then the limit of stack contents is $\bot_1.u^\tla.\#^\om$.
\nl The $\om$-word $\bot_1.u^\tla.\#^\om$
is accepted by $\mathcal{A}_2$ iff $u^\tla \in L$.
\nl Thus the  configuration $(q, \bot.u)$ is
a winning position for Eve in the induced pushdown game,
equipped with the winning condition
$\Omega_{\mathcal{A}_1 \rhd \mathcal{A}_2}$, if and only if
$u \in L^\sim$.
\ep

\hs We can now state the following result which  follows
directly from Lemmas \ref{lemna} and \ref{lemna2}.

\begin{Pro}\label{prows}
There exists a pushdown process $\mathcal{P}=(Q, \Ga,  \bot,  \delta)$, a
partition
$Q = Q_E \cup Q_A$,  two deterministic  pushdown automata $\mathcal{A}_1,
\mathcal{A}_2$,
and a state  $q\in Q$ such that, in the induced pushdown game
equipped with the winning condition
$\Omega_{\mathcal{A}_1 \rhd \mathcal{A}_2}$,
   the  set $\{u \mid (q, u) \in W_E \}$ is a
{\it non-deterministic} non ambiguous context free language.
\end{Pro}

\proo Let  $L$ be  the  language
$\{a^n.b^n \mid  n\geq 1\}$.
 The language $L$ is a deterministic  context free language, 
  thus  by Lemma \ref{lemna2} 
there
exists a pushdown process $\mathcal{P}=(Q, \Ga,  \bot,  \delta)$, a
partition
$Q = Q_E \cup Q_A$,  two deterministic  pushdown automata $\mathcal{A}_1,
\mathcal{A}_2$,
and a state  $q\in Q$ such that, in the induced pushdown game
equipped with the winning condition
$\Omega_{\mathcal{A}_1 \rhd \mathcal{A}_2}$,
one has  $\{u \mid (q, u) \in W_E \} = L^\sim$.
But by Lemma \ref{lemna}  $L^\sim$ is a non ambiguous
context free language
which can not be accepted by any {\it deterministic
pushdown automaton}.
\ep

\begin{Rem}
In the  pushdown game given in the proof of Lemma \ref{lemna2},
there are some  plays which are not  infinite. However it is easy
to find a pushdown game with the same winning set for Eve but in which all
plays are infinite.
The same remark will hold for pushdown games given in the proofs of the two
following
propositions.

\end{Rem}

\noi We are now going to show that the set of winning positions for Eve
can also be an inherently  ambiguous context free language.
Recall that it is well known that
the language $V=\{a^n.b^m.c^p \mid n, m, p \geq 1 \mbox{ and } ( n=m
\mbox{ or } m=p )\}$
is an inherently  ambiguous context free language,
\cite{ABB96,HopcroftMotwaniUllman2001}.

\begin{Pro}\label{inamb}
There exists a pushdown process $\mathcal{P}=(Q, \Ga,  \bot,  \delta)$, a
partition
$Q = Q_E \cup Q_A$,  two deterministic  pushdown automata $\mathcal{A}_1,
\mathcal{A}_2$,
and a state  $q\in Q$ such that, in the induced pushdown game
equipped with the winning condition
$\Omega_{\mathcal{A}_1 \rhd \mathcal{A}_2}$,
   the  set $\{u \mid (q, u) \in W_E \}$ is an
inherently  ambiguous context free language.
\end{Pro}

\proo Let $\mathcal{P}=(\{q, q', q'', p\}, \Ga=\{\bot, a, b, c, \#\},  \bot,
\delta)$
be a pushdown process
where $\delta$ is defined by: $\{ pop(q'), skip(q'') \} \subseteq  \delta(q,
c)$,
$pop(q') \in \delta(q', c)$, $push(p, \#) \in \delta(q', b)$,
$push(p, \#) \in \delta(q'', c)$, and  $push(p, \#) \in \delta(p, \#)$.
\nl We set $Q_E=\{q\}$ and $Q_A=\{q', q'', p\}$.
\nl Consider now an infinite play from a given configuration $(q, \bot.u)$,
for
$u\in \{a, b, c, \#\}^\star$. The topmost stack letter of this initial
configuration
must be a letter $c$.  Then at most two cases may happen.
\begin{enumerate}
\ite In the first one are pushed infinitely many letters $\#$ on the top of
the stack.
In this play
the stack is strictly unbounded and the limit of the stack contents is
$\bot.u.\#^\om$.
\ite  In the second case the letter $c$ is popped and all next letters $c$
are popped from the
top of the stack until some letter $b$ is on the top of the stack. From this
moment
infinitely many letters $\#$ are pushed in the stack. Then
the stack is strictly unbounded and the limit of the stack contents is
$\bot.u'.b.\#^\om$
if $u=u'.b.c^k$ for some integer $k>0$. Notice that this second case
can only occur if $u$ is in the form $u=u'.b.c^k$ for some integer $k>0$.
\end{enumerate}

\noi The deterministic pushdown automaton $\mathcal{A}_1$ reads words over
the alphabet
$\{\bot, a, b, c,   \#\}$ and its stack alphabet is  $\Ga_1=\{\bot_1, a, b,
\#\}$.
\nl It is easy to ensure that the stack of $\mathcal{A}_1$ is not strictly
unbounded during
the reading of an $\om$-word which is not in
$W = \bot.a^\om \cup \bot.a^+.b^\om \cup  \bot.a^+.b^+.\#^\om
\cup  \bot.a^+.b^+.c^\om
\cup \bot.a^+.b^+.c^+.\#^\om$.
\nl Consider now the reading by $\mathcal{A}_1$ of an $\om$-word which is in
$W$. After
having read the bottom symbol $\bot$, the stack content of $\mathcal{A}_1$
is still $\bot_1$.
Then it pushes a letter $a$ or $b$ each time it reads the corresponding
letter $a$ or $b$.
\nl Then when  $\mathcal{A}_1$ reads an $\om$-word in the form $\bot.a^\om$
(respectively,
$\bot.a^n.b^\om$ for $n\geq 1$) then its  stack is  strictly unbounded and
the limit of stack contents is
$\bot_1.a^\om$ (respectively, $\bot_1.a^n.b^\om$).
\nl If now $\mathcal{A}_1$ reads  letters $\#$ then it pushes them in the
stack. In this case
the input word is in the form $\bot.a^n.b^m.\#^\om$, and the limit of stack
contents of
$\mathcal{A}_1$ reading this $\om$-word is $\bot_1.a^n.b^m.\#^\om$.
\nl If $\mathcal{A}_1$ reads  some letters $c$ after an initial segment in
the form
$\bot.a^n.b^m$ then it pops a letter $b$ for each letter $c$ read.
\nl If the number of $c$ is equal to the number of $b$ of the input word,
then after having
read the segment $\bot.a^n.b^m.c^m$ of the input word the stack content of
$\mathcal{A}_1$
is simply $\bot_1.a^n$.
Next $\mathcal{A}_1$ reads the final segment  $\#^\om$
and it pushes it in the stack. So  the limit of stack contents of
$\mathcal{A}_1$ reading the input $\om$-word $\bot.a^n.b^m.c^m.\#^\om$
is in the form
 $\bot_1.a^n.\#^\om$
\nl If the number of $c$ is not equal to the number of $b$ of the input word
(the number
of $c$ being finite or infinite), then, once this has been checked,
the stack content remains unchanged so the stack will not be strictly
unbounded.

\hs One can  define a deterministic pushdown automaton $\mathcal{A}_2$,
equipped
with a parity acceptance condition, which accepts the $\om$-language
$\{\bot_1.a^n.b^n.\#^\om \mid n\geq 1\} \cup \{\bot_1.a^n.\#^\om \mid n\geq
1\} $.

\hs We are now going to determine the winning positions $(q, \bot.u)$ of Eve
in
the induced  pushdown game equipped with the winning condition
$\Omega_{\mathcal{A}_1 \rhd \mathcal{A}_2}$.

\hs Let  $(q, \bot.u)$ be
a given configuration  of the pushdown process $\mathcal{P}$
for some $u \in \{a, b, c, \#\}^\star$, the last letter of $u$ being  $c$.
There are one or two infinite plays starting from this position.
When there are two such plays, they depend on the first choice of
Eve and  the position  $(q, \bot.u)$
  is a  winning position for
Eve iff one of the two possible infinite plays is winning for her.
\nl
In the first play  the stack is strictly unbounded and
the limit of the stack contents is $\bot.u.\#^\om$.
\nl There is a second play if  $u=u'.b.c^k$ for some integer $k>0$.
Then in this play the stack is strictly unbounded and
the limit of the stack contents is $\bot.u'.b.\#^\om$.

\hs  When $\mathcal{A}_1$ reads the $\om$-word $\bot.u.\#^\om$,   its stack
is
strictly unbounded iff  $u$ is in the form  $a^n.b^m.c^m$  for some $n, m
\geq 1$
(the number of $c$ and of $b$ in $u$ are equal).
Then the limit of stack contents is $\bot_1.a^n.\#^\om$ and it is in
$L(\mathcal{A}_2)$.
So  $\bot.u.\#^\om \in L(\mathcal{A}_1 \rhd \mathcal{A}_2)$.

\hs If  $u=u'.b.c^k$ for some integer $k>0$ and $\mathcal{A}_1$ reads the
$\om$-word
$\bot.u'.b.\#^\om$ then the stack of $\mathcal{A}_1$ is
strictly unbounded  iff $u'$ is in the form $a^n.b^{m-1}$ for some $n, m\geq
1$.
In this case the limit of stack contents is $\bot_1.a^n.b^m.\#^\om$ and it
is accepted by
$\mathcal{A}_2$ iff $n=m\geq 1$.

\hs Thus the  configuration $(q, \bot.u)$ is
a winning position for Eve, with the winning condition
$\Omega_{\mathcal{A}_1 \rhd \mathcal{A}_2}$, if and only if
$u$ is in the inherently  ambiguous context free language
$V=\{a^n.b^m.c^p \mid n, m, p \geq 1 \mbox{ and } ( n=m \mbox{ or }
m=p )\}$.
\ep

\begin{Pro}
There exists a pushdown process $\mathcal{P}=(Q, \Ga,  \bot,  \delta)$, a
partition
$Q = Q_E \cup Q_A$,  two deterministic  pushdown automata $\mathcal{A}_1,
\mathcal{A}_2$,
and a state  $q\in Q$ such that, in the induced pushdown game
equipped with the winning condition
$\Omega_{\mathcal{A}_1 \rhd \mathcal{A}_2}$,
   the  set $\{u \mid (q, u) \in W_E \}$ is a
non context free language.
\end{Pro}

\proo We define the  pushdown process $\mathcal{P}=(Q, \Ga,  \bot,  \delta)$
as in the proof of preceding Proposition \ref{inamb} except that we set this
time
$Q_A=\{q\}$ and $Q_E=\{q', q'', p\}$. The two deterministic
 pushdown automata $\mathcal{A}_1,  \mathcal{A}_2$, are also defined in the
same way.

\hs Consider now a configuration in the form $(q, \bot.a^n.b^m.c^p)$ for
some
integers $n, m, p \geq 1$. There are two infinite plays starting from this
configuration
but they depend this time on the first choice of {\it the second player
Adam}.
\nl The position  $(q, \bot.a^n.b^m.c^p)$ is winning for Eve iff these {\it
two infinite plays}
are won by her. This implies that $n=m$ {\bf and} $m=p$.
\nl Thus it holds that
$$\{ u \mid (q,u)\in W_E \} ~\cap ~\bot.a^+.b^+.c^+ = \bot.\{a^n.b^n.c^n
\mid n\geq 1 \}$$
\noi This language is not context free because of the well known non context
freeness of
the language $\{a^n.b^n.c^n \mid n\geq 1 \}$
\cite{ABB96,HopcroftMotwaniUllman2001}.
\nl This implies that the set $\{ u \mid (q,u)\in W_E \}$ itself is not
context free. Indeed
otherwise its intersection with the rational language  $\bot.a^+.b^+.c^+$
would be
context free because the class $CFL$ is closed under intersection with
rational languages.
\ep

\hs {\bf  Acknowledgements.} Thanks to the anonymous referee
for useful comments on a preliminary version of this paper.

\end{document}